\documentstyle[12pt,lathuile,epsfig]{article}
\def\circa#1{\,\raise.3ex\hbox{$#1$\kern-.75em\lower1ex\hbox{$\sim$}}\,}

\newcommand  \f  \varphi

\newcommand{\be}{\begin{equation}}
\newcommand{\ee}{\end{equation}}
\newcommand{\ben}{\begin{displaymath}}
\newcommand{\een}{\end{displaymath}}
\newcommand{\ba}{\begin{eqnarray}}
\newcommand{\ea}{\end{eqnarray}}
\newcommand{\ban}{\begin{eqnarray*}}
\newcommand{\ean}{\end{eqnarray*}}

\begin{document}
\title{ TeV scale weak interactions: 
inclusive vs. exclusive observables}

\author{Paolo Ciafaloni       \\
{\em INFN - Sezione di Lecce} \\
{\em E-mail: paolo.ciafaloni@le.infn.it}
}
\maketitle
\baselineskip=14.5pt
\begin{abstract}
Hard processes at the TeV scale exhibit enhanced (double log) 
EW  corrections, that need resummation in view of the high level of precision
of Next Linear Colliders. The fact that the weak sector is spontaneously
broken causes some peculiarities with respect to unbroken theories
like QCD.  
For observables that are 
exclusive with respect to  $W,Z$ emission, some peculiar
technical problems have yet to be solved.
Surprisingly, double logarithmic enhancements  are present 
even for inclusive observables like $e^+e^-\to$ hadrons,
leading to violation of the Bloch-Nordsieck theorem.
The last effect is particularly important, producing weak effects
that in some cases compete in magnitude with the strong ones.
\end{abstract}
\baselineskip=17pt
\newpage



\section{Introduction}

In last two years, starting from the observation made in\cite{cc1},
it has become clear that the bulk of radiative electroweak (EW)
corrections at the
TeV scale is given by logarithms of infrared origin, also called
Sudakov logarithms\cite{suda}. 
Such logarithms occur because at energies much larger than the
EW scale $M_Z\approx M_W\equiv M$, the latter acts as a cutoff for the
collinear and infrared (IR) divergences that would be present in the vanishing
$M$ limit.

Due to this (double) logarithmic enhancement, EW corrections
become pretty big at the TeV scale, producing corrections 
of the order of 10 $\%$
to cross sections. 
Then, resummation becomes necessary and has been addressed
for instance in\cite{kuhn,cc2,fadin}
giving rise to some still unresolved controversy. Partly because of this
controversy, one can state that Standard Model EW 
corrections are not under control at the 
1 $\%$ 
level at the TeV scale\cite{mio}. 
Moreover, this kind of corrections is ubiquitous,
being present also in inclusive quantities as has 
been noticed in\cite{3p,last}.

There are various aspects of interest in these  issues. First,
one of the main goals of TeV scale accelerators like 
Next  Linear Colliders\cite{NLC} 
will be to explore possibilities for New Physics with a high level
of precision.
However, it  is clear that
no serious limit on, say, 
 anomalous gauge couplings can be given and,
quite  in general, no signal of New Physics can be established until the
abovementioned questions are resolved and established on a firm theoretical
ground. Second,  in particular papers\cite{3p,last}  make it clear
that the
IR dynamics of a broken theory like the SM electroweak sector is
poorly known at the moment and might give rise to surprises; the
Block-Nordsieck theorem is violated for instance. Therefore, in my opinion
studying and testing this subject can be {\sl in itself} 
a reason of interest for NLCs experiments.

In the following I consider processes of the kind 
2 fermions $\to$ 2 fermions+X, characterized by a single hard scale, typically
the c.m energy $\sqrt{s}$, much greater than the EW symmetry breaking scale
$M$. Here X represents emitted soft 
 weak bosons of energy $\omega\ll \sqrt{s}$.
All other hard scales are are of the same order,
namely $|s|\sim |t|\sim |u|\gg M$. Two cases are taken in exam:
observables that are 
{\sl inclusive} with respect to $W,Z$ emission (Par. 3) and observables
that are {\sl exclusive}   with respect to $W,Z$ emission (Par. 2), but still 
include soft photon radiation up to a given experimental resolution 
 $\lambda$. One can summarize the present situation like this:
\begin{itemize}
\item
The study of exclusive EW form factors at the TeV scale is a challenging one
and results are still controversial; a complete two loop calculation could
help in solving some of the open questions.
\item
inclusive observables are characterized by unsuppressed double logarithmic
corrections of IR origin, leading to violation of the BN theorem and to weak
corrections that can be of the order of the strong ones.
\end{itemize}

\section{Exclusive observables}
Exclusive observables are characterized by 3 energy scales: the c.m. energy
$\sqrt{s}$, the symmetry breaking scale $M_W\sim M_Z\equiv M$ and the
infrared cutoff $\lambda_{IR}$\footnote{fermions can be taken to be massless
as long as $\lambda\gg m_f$; the special case of the top quark, 
requiring a heavy mass
cutoff, was considered in\cite{lecce}, but leads to no important differences
at the double log level.}. 
One must always be inclusive with respect to
radiated photons up to a certain energy/angle resolution $\lambda$, which
amounts to making the substitution $\lambda_{IR}\to\lambda$. 
A typical example is $e^+e^-\to \mu^+\mu^-+X$, where $X$ is a
soft photon.
The presence of 3
scales is a major difference with unbroken theories like QCD where the
analogous problem is characterized by a single expansion parameter 
$\log\frac{\sqrt{s}}{\lambda}$; here, there are two expansion
parameters $L\equiv\log\frac{\sqrt{s}}{M}$ and  
$l\equiv\log\frac{M}{\lambda}$ (see\cite{cc2}). 
Since right fermions do not carry
non abelian charges,  
the interesting case to consider is
the one with left (L) fermions on the external legs.  One limit in which one
already knows what should happen is the SU(2)$\otimes$ U(1)
symmetric limit, i.e. $\sqrt{s}\gg M,\lambda$. 
In this regime the resummed matrix element
is given in terms of the Born one by the following expression
\be
{\bf M}^{L}=\exp[-\sum_i\frac{C^F_i}{2}
\log^2\frac{s}{M^2}]\;{\bf M}_0^L\qquad
C^F_i=g'^2y_i^2+g^2 \frac{3}{4}
\ee
involving the sum of the Casimir $C^F_i$ in the fundamental representation
over all external legs $i$. However, if the energy $\sqrt{s}$ is not extremely
large, the situation is more complicated, due
to the presence of the EW symmetry breaking scale $M$. The
main point is that there is a separation of scales (see Fig. 1) such that 
QED soft effects are present below $M$, while the full EW contributions
($\gamma,Z,W$) has to be taken into account above $M$. As a byproduct of this
picture, it has been shown in\cite{cc2} that taking into account QED soft
effects {\sl separately} as was customary for LEP\cite{LEP,YFS}, 
is not anymore correct at the TeV scale.

As already said, controversial results are present 
in the literature\cite{kuhn,cc2,fadin}, 
indicating that symmetry breaking makes life harder for
this kind of problem. The problem arises, namely, about the details of the
scale separation and about the role of symmetry breaking when the three scales
are relatively close to each other. The fact is that when one considers
resummation to all orders, one is forced to make a priori assumptions
for the calculation to be possible.
{\sl A complete two-loop calculation} without any a priori assumption
would shed light on this controversy. A sort of ``minimal calculation''
sufficient for this purpose could be the process $Z'\to f\bar{f}$,
considered in\cite{cc2}.

\section{Inclusive observables}
The unique, and surprising, feature of EW interactions with respect
to nonabelian unbroken theories like QCD, is the violation
of the Bloch-Nordsieck (BN) theorem\cite{bl}.
In abelian theories like QED, although radiative corrections are IR divergent
in general, one recovers a finite result by summation over all possible
degenerate finale states; this is the essence of the BN theorem. In particular
then, if one regularizes IR divergences by introducing a cutoff
$\lambda_{IR}$, inclusive observables do not depend at all on this cutoff. 
In principle, the BN theorem is violated in {\sl any} nonabelian theory, like
QCD for instance, since to recover an IR finite result one should 
sum also over {\sl initial} degenerate states\cite{KLN}, which is however
unphysical in general. In QCD one is saved at the bottom line due to color
confinement: the initial states are color singlets, and averaging over
initial color produces cancellation of the leading IR divergences\cite{dft}.

The crucial observation of\cite{3p,last} is that the situation with EW
interactions is very different from QCD. In fact 
the colliders initial states   ($e^-,p,\nu...$)
carry in general a nonabelian weak isospin charge, and averaging over the
initial isospin makes no sense from an experimental point of view. This means
that the BN theorem is violated, and that even inclusive observables retain
the leading (double log) dependence on the IR cutoff $M$. The effect is quite
dramatic for the typical case of $e^+e^-\to $ hadrons: while QCD corrections
are perturbative, and therefore
 almost energy independent, EW effects steadily grow
with energy and get as big  as the strong ones (see Fig. 2): a sort of early
unification!

What happens here is that the nonabelian gauge 
structure of the theory and the breaking of the gauge symmetry itself
conspire to produce a nontrivial result. Symmetry breaking is crucial since,
besides providing the physical cutoff $M$, it gives a finite range to weak
interactions allowing for asymptotic initial states that are ``bare''
nonabelian charges like $e^-_L$; the analogous situation in QCD is forbidden.

\begin{figure}[htb]\setlength{\unitlength}{1cm}
\begin{picture}(10,7)
\put(3.7,6.2){QED {$\quad \lambda<w<M$}}
\put(3.7,4.1){EW {$\quad M<w<\sqrt{s}$}}
\put(3,0){\epsfig{file=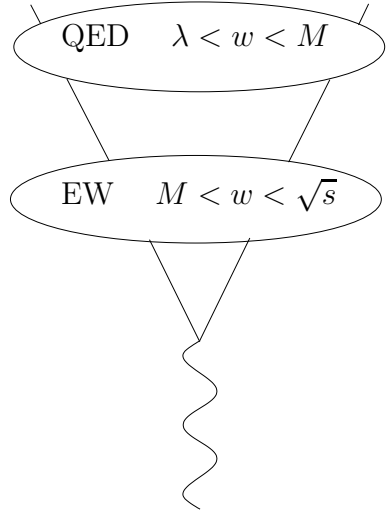,height=6.8cm}}
\end{picture}
\caption{Separation of scales for EW IR effects at the TeV scale. The process
considered is $Z'\to f\bar{f}$ (see\cite{cc2} for details).}
\end{figure}
\begin{figure}[b]\setlength{\unitlength}{1cm}
\begin{picture}(12,8)
\put(3,0){\epsfig{file=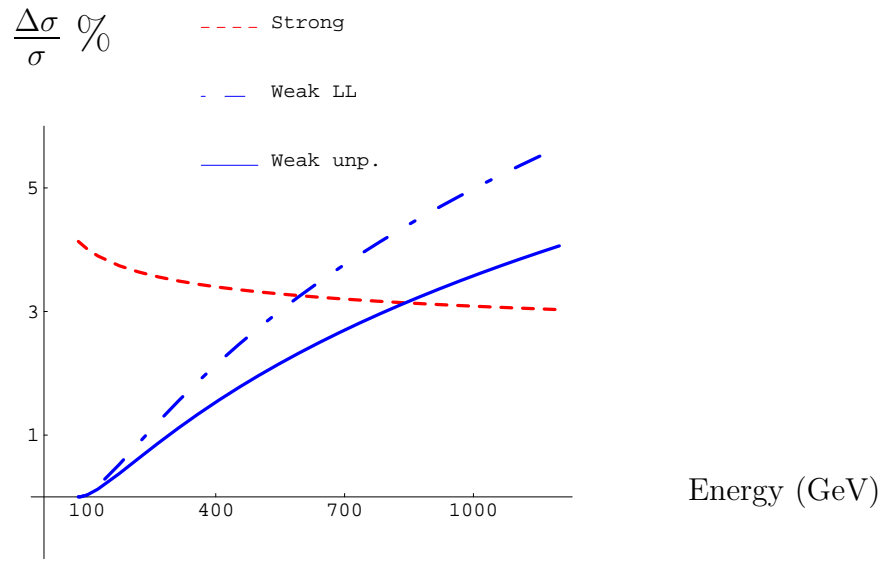,height=8cm}}
\put(3,7){\Large $\frac{\Delta\sigma}{\sigma}$ \%}
\put(12,1){Energy (GeV) }
\end{picture}
\caption{\label{had} Resummed double log EW corrections to $e^+e^-\to$ hadrons 
and strong corrections (dashed line) up to 3 loops. 
The dash-dotted line is
for a LL polarized beam, while the continuous line is for an 
unpolarized beam.}
\end{figure}

\end{document}